\documentclass{webofc}
\usepackage[varg]{txfonts}   

\def\lsim{\raise0.3ex\hbox{$<$\kern-0.75em\raise-1.1ex\hbox{$\sim$}}}
\def\gsim{\raise0.3ex\hbox{$>$\kern-0.75em\raise-1.1ex\hbox{$\sim$}}}

\begin{document}
\title{Heavy-flavor transport and hadronization in a small fireball}
%
%

\author{\firstname{Andrea} \lastname{Beraudo}\inst{1}\fnsep\thanks{\email{beraudo@to.infn.it}} \and
        \firstname{Arturo} \lastname{De Pace}\inst{1}\fnsep \and
        \firstname{Daniel} \lastname{Pablos}\inst{1,2,3}\fnsep \and
        \firstname{Francesco} \lastname{Prino}\inst{1}\fnsep \and
        \firstname{Marco} \lastname{Monteno}\inst{1}\fnsep \and
         \firstname{Marzia} \lastname{Nardi}\inst{1}}

\institute{INFN - Sezione di Torino, via Pietro Giuria 1, IT-10125 Torino
\and
           Departamento de F\'isica, Universidad de Oviedo, Avda. Federico Garc\'ia Lorca 18, ES-33007 Oviedo 
\and
           Instituto Universitario de Ciencias y Tecnolog\'ias Espaciales de Asturias (ICTEA), Calle de la Independencia 13, ES-33004 Oviedo
          }

\abstract{We study heavy-flavor hadron production in high-energy pp collisions, assuming the formation of a small, deconfined and expanding fireball where charm quarks can undergo rescattering and hadronization. We adopt the same in-medium hadronization mechanism developed for heavy-ion collisions, which involves Local Color-Neutralization (LCN) through recombination of charm quarks with nearby opposite color charges from the background fireball. Diquark excitations in the hot medium favor the formation of charmed baryons. The recombination process, involving closely aligned partons from the same fluid cell, effectively transfers the collective flow of the system to the final charmed hadrons. This framework can qualitatively reproduce the observed experimental findings in heavy-flavor particle-yield ratios, $p_T$-spectra and elliptic-flow coefficients. Our results provide new, complementary support to the idea that the collective phenomena observed in small systems have the same origin as those observed in heavy-ion collisions.}
\maketitle
\section{Introduction}
\label{sec:intro}
The recently observed strong enhancement of the charmed baryon-to-meson ratio in pp collisions~\cite{ALICE:2020wfu}, incompatible with hadronization models tuned to reproduce $e^+e^-$ data, is on the contrary consistent with the results obtained in heavy-ion collisions. 
Accordingly, we propose that the same mechanism of heavy-flavor (HF) hadron production at work in heavy-ion collisions occurs in the pp case, assuming that also in proton-proton collisions a small deconfined fireball, with a hydrodynamic expansion driven by pressure gradients, is formed. Similar ideas were proposed in Refs.~\cite{Minissale:2020bif,Song:2018tpv}.
In our approach the hot medium affects the stochastic propagation, modeled through a relativistic Langevin equation, of the heavy quarks before hadronization and acts as a reservoir of color charges with which they can undergo recombination when reaching a fluid cell around the QCD hadronization temperature $T_H$. More details on the implementation of our model can be found in Refs.~\cite{Beraudo:2022dpz,Beraudo:2023nlq} referring to the AA and pp case, respectively.
\section{Local Color-Neutralization mechanism}
\label{sec:LCN}
\begin{figure}[h]
\centering
\includegraphics[clip,height=4cm]{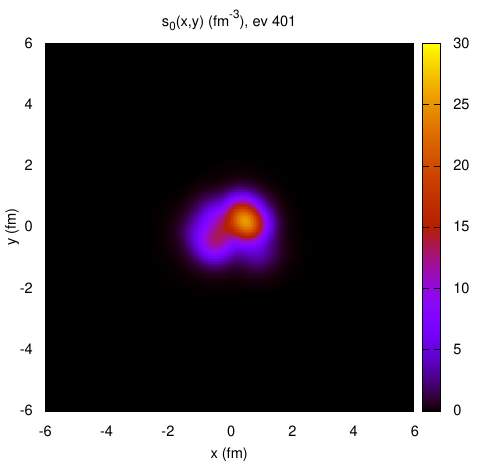}
\hspace{1cm}
\includegraphics[clip,height=4cm]{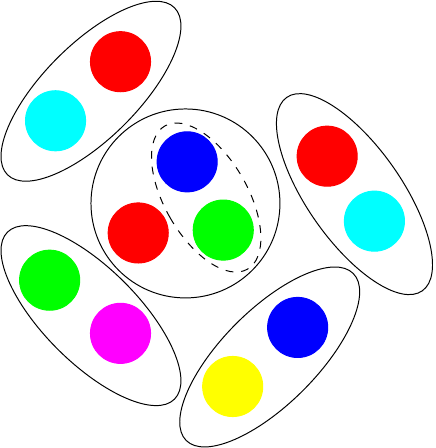}
\caption{Left panel: the initial entropy-density profile for a given pp event belonging to the minimum-bias sample. Right panel: a carton of the LCN process, in which (anti-)quarks are recombined with the closest opposite color-charge, giving rise to singlet clusters.}
\label{fig:cartoon}       
\end{figure}
In Ref.~\cite{Beraudo:2023nlq} we showed that, if a small fireball is formed also in pp collisions (see left panel of Fig.~\ref{fig:cartoon}), heavy-quark pairs tend to be produced in the hot spots of the highest multiplicity events. Hence, the presence of a dense partonic environment -- here described as a deconfined plasma close to local thermal equilibrium -- affects their propagation and eventual hadronization. Here we focus on this last stage, modeled as a Local Color Neutralization (LCN) process in which, around the QCD crossover temperature, a heavy quark undergoes recombination with the \emph{closest} opposite color charge (see right panel of Fig.~\ref{fig:cartoon}); the latter can be either an antiquark or a diquark, assumed to be present in the fireball with a thermal abundance dictated by their mass. A color-singlet cluster is thus formed, which can carry baryon number $B=0,\pm 1$. Both the screening of color interaction in a dense environment and the minimization of the energy stored in the confining potential justify the locality of the color-blenching process, which has to involve nearby partons. This introduces a strong Space-Momentum Correlation (SMC), since particles from the same fluid cell tend to share a common collective velocity. The formed clusters eventually undergo a decay into the final hadrons. Such a $2\to 1\to N$ process ensures exact four-momentum conservation, at variance with coalescence approaches. Since SMC favors the recombination of collinear partons, the formed clusters typically have a low invariant mass. Hence they are assumed to undergo an isotropic 2-body decay in their local rest-frame into a charmed hadron and a soft particle (a pion in most cases). Heavier clusters ($M\gsim 4$ GeV) are instead fragmented as Lund strings, employing PYTHIA 6.4~\cite{Sjostrand:2006za}. More details on our LCN model ca be found in~\cite{Beraudo:2022dpz}.
\section{Results}
Here we show how our in-medium transport+hadronization model, originally developed to describe nuclear collisions, once applied to pp collisions at LHC energies (our simulations refer to $\sqrt{s}\!=\!5.02$ TeV) can provide a consistent picture of HF hadron production also in these systems, without changing any of its parameters. 
\begin{figure}[h]
\centering
\includegraphics[width=0.9\textwidth,clip]{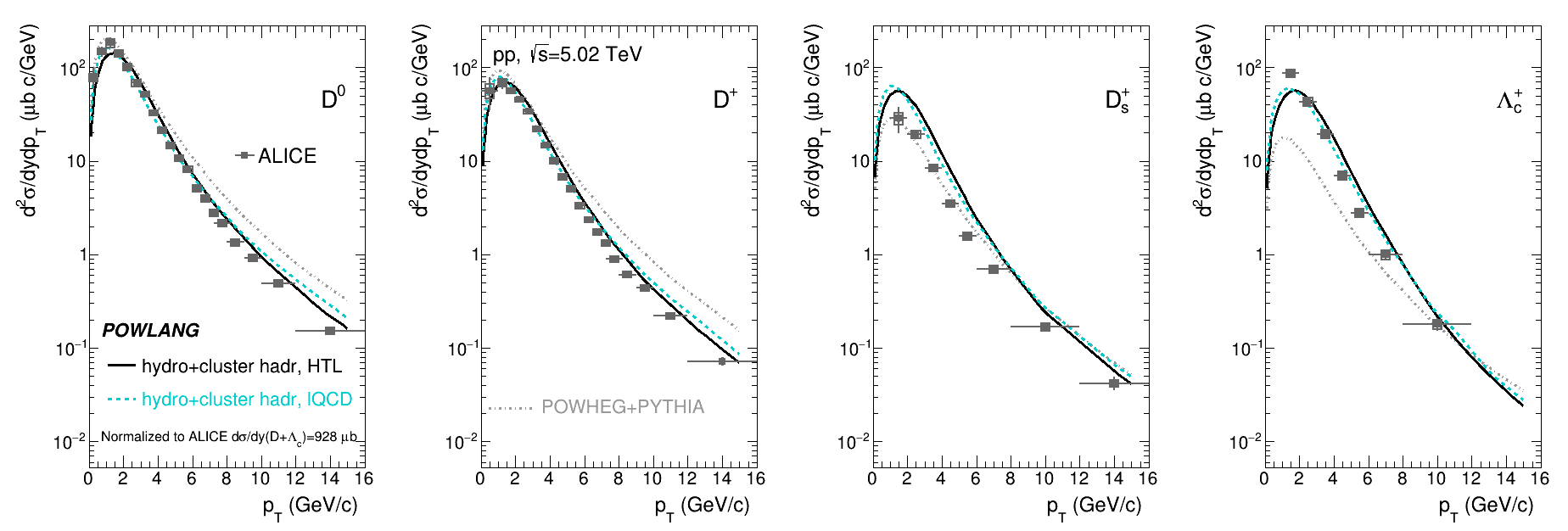}
\caption{Charmed hadron $p_T$-distributions in pp collisions at $\sqrt{s}\!=\!5.02$ TeV normalized to the experimental $D+\Lambda_c^+$ cross-section. Results obtained with POWHEG-BOX standalone (dotted grey curves) or supplemented with an in-medium transport+hadronization stage with HTL (continuous black curves) and lattice-QCD (dashed cyan curves) are compared to ALICE data~\cite{ALICE:2020wfu,ALICE:2021mgk}.}
\label{fig:spectra}       
\end{figure}
We start considering the charmed-hadron $p_T$-distributions, plotted in Fig.~\ref{fig:spectra}. As one can see, the slope of the spectra cannot be reproduced by a standard pQCD event generator (dotted grey curves) like POWHEG-BOX~\cite{Alioli:2010xd}, but a better agreement with the experimental data can be obtained including medium corrections to the transport and hadronization of heavy quarks.

\begin{figure}[h]
\centering
\includegraphics[width=0.9\textwidth,clip]{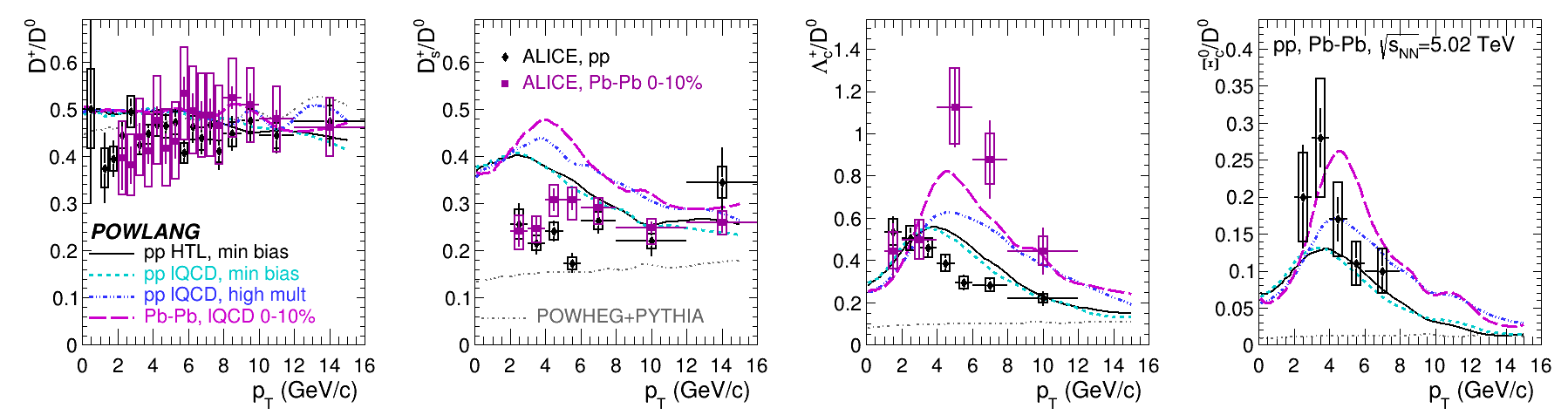}
\caption{Charmed-hadron yield ratios as a function of $p_T$ at $\sqrt{s_{\rm NN}}\!=\!5.02$ TeV. Predictions including in-medium transport+hadronization in minimum-bias and high-multiplicity pp collisions and in central PbPb collisions are compared to ALICE data~\cite{ALICE:2020wfu,ALICE:2021mgk,ALICE:2021rxa,ALICE:2021bib,ALICE:2021psx}. The enhanced baryon-to-meson ratio and the shift of its peak in denser systems is qualitatively well reproduced. Also shown are the pp predictions obtained with POWHEG+PYTHIA standalone, undershooting charmed baryon production.}
\label{fig:ratio}       
\end{figure}
Moving then to the charmed-hadron yield ratios, one can see that the astonishing enhanced baryon-to-meson ratio observed both in AA and in pp collisions is qualitatively well described our LCN process occurring at the end of the transport stage. Moving from minimum-bias to high-multiplicity pp collisions and, finally, to central Pb-Pb collisions the ratios of the integrated charmed-hadron yields remain pretty constant; however, when plotted as a function of $p_T$, the peaks in the $\Lambda_c^+/D^0$ and $\Xi_c^0/D^0$ ratios move to higher momenta, due to the larger radial flow of the light diquarks involved in the recombination process.

\begin{figure}[h]
\centering
\includegraphics[clip,height=4cm]{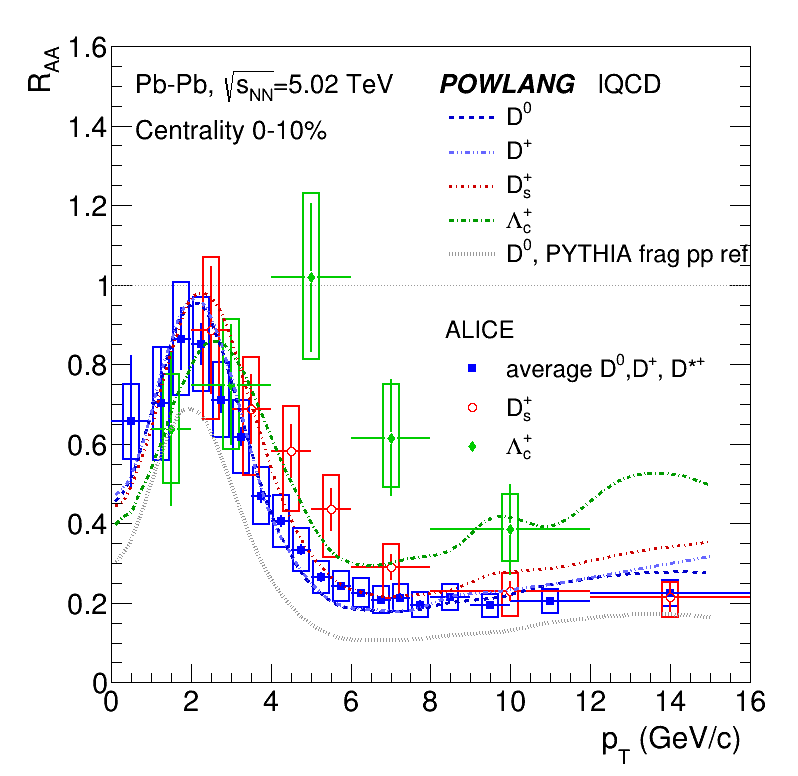}
\hspace{1cm}
\includegraphics[clip,height=4cm]{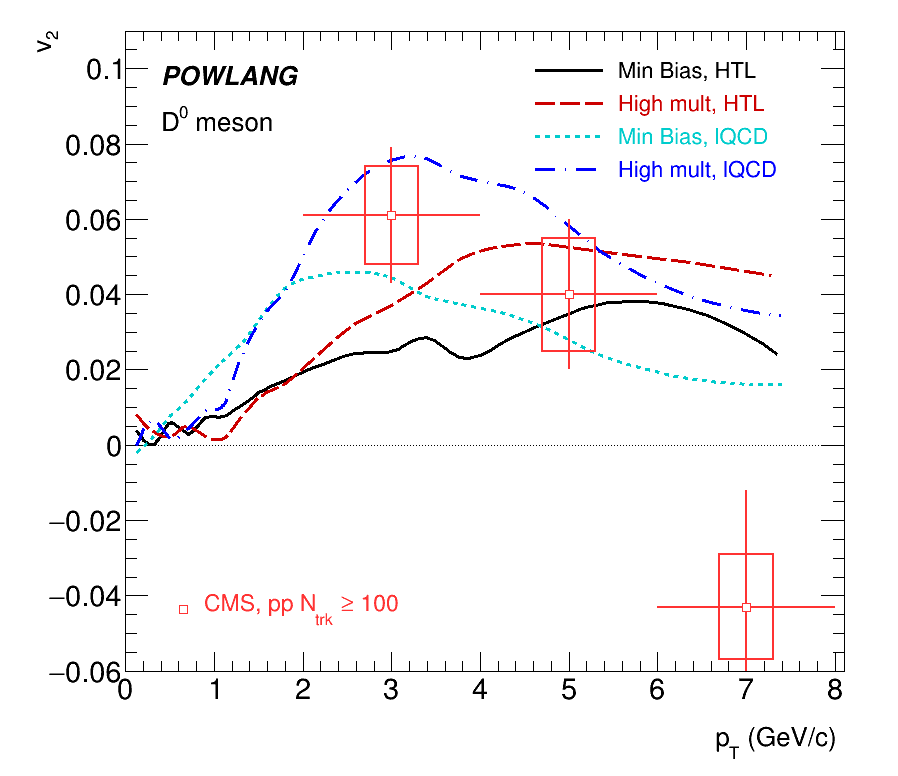}
\caption{Left panel: effect on the charmed-hadron $R_{\rm AA}$ of including in-medium transport+hadronization in the pp benchmark. Theory curves are compared to ALICE data~\cite{ALICE:2021rxa,ALICE:2021bib}; also shown is the $D^0$ result with no medium effect in the pp collisions. Right panel: charmed-hadron elliptic-flow coefficient in minimum-bias and high-multiplicity pp collisions at $\sqrt{s}\!=\!5.02$ TeV. Our predictions are compared to CMS results for high-multiplicity pp collisions at $\sqrt{s}\!=\!13$ TeV~\cite{CMS:2020qul}.}
\label{fig:RAA-v2}       
\end{figure}
Our study is relevant to correctly quantify medium effects in heavy-ion collisions, where the pp benchmark enters in defining the nuclear modification factor $R_{\rm AA}(p_T)\!\propto\!(dN/dp_T)_{\rm AA}/(dN/dp_T)_{\rm pp}$. 
As one can see in the left panel of Fig.~\ref{fig:RAA-v2}, the inclusion of medium effects in pp collisions is necessary to correctly reproduce the location and magnitude of the radial-flow peak (i.e. the reshuffling of the particle momenta, moving from low to moderate $p_T$) and to obtain a species dependence of the results with the same qualitative trend of the experimental data.

Finally, we also provide predictions for the elliptic-flow coefficients $v_2$, displayed in the right panel of Fig.~\ref{fig:RAA-v2} for minimum-bias and high-multiplicity pp collisions at $\sqrt{s}\!=\!5.02$ TeV and compared to CMS results at $\sqrt{s}\!=\!13$ TeV~\cite{CMS:2020qul}.

\section{Discussion}
The same Local Color-Neutralization (LCN) model developed to describe medium-modification of HF hadronization in AA collisions has been extended to model charmed-hadron production in the pp case.
Our major finding is that our LCN model coupled to transport calculations in a small fireball provides a consistent comprehensive description of several HF observables: the shape of the $p_T$-distributions, the enhanced baryon-to-meson ratio, the charmed-hadron $R_{\rm AA}$ and the non-vanishing $v_2$ coefficient.
Our results provide independent, strong indications that the collective phenomena observed in small systems may have the same origin as those measured in heavy-ion collisions.

\section*{Acknowledgements}
D.P. has received funding from the European Union’s Horizon 2020 research and innovation program under the Marie Sklodowska-Curie grant agreement No. 754496. 

\bibliography{QM-beraudo}

\end{document}